\documentclass[twocolumn,twoside,slac_two]{revtex4}
\usepackage{graphics,graphicx}
\usepackage{fancyhdr}
\newcommand{\Msun}{\>{\rm M_{\odot}}}

\newcommand\apj{ApJ}
\newcommand\apjl{ApJ}

\newcommand\aj{AJ}

\newcommand\mnras{MNRAS}

\begin{document}

\title{Observational Signature of Tidal Disruption of a Star by a Massive Black Hole}
\author{T. Bogdanovi\'c, M. Eracleous, S. Sigurdsson, P. Laguna}
\affiliation{Pennsylvania State University, PA 16802, USA}
\author{S. Mahadevan}
\affiliation{University of Florida, FL 32611, USA}

\begin{abstract}
We have modeled the time-variable profiles of the H$\alpha$ emission
from the nonaxisymmetric disk and debris tail created in the tidal
disruption of a solar-type star by a $10^{6}\Msun$ black
hole. Two tidal disruption events were simulated using a three
dimensional relativistic smoothed particle hydrodynamic code to
describe the early evolution of the debris during the first 50-90
days. We have calculated the physical conditions and radiative
processes in the debris using the photoionization code CLOUDY. We
model the emission-line profiles in the period immediately after the
accretion rate onto the black hole becomes significant. We find that
the line profiles at these very early stages of the evolution of the
postdisruption debris do not resemble the double-peaked profiles
expected from a rotating disk, since the debris has not yet settled
into such a stable structure. As a result of the uneven distribution
of the debris and the existence of a ``tidal tail'' (the stream of
returning debris), the line profiles depend sensitively on the
orientation of the tail relative to the line of sight. Moreover, the
predicted line profiles vary on fairly short time scales (of order
hours to days). Given the accretion rate onto the black hole we also
model the H$\alpha$ light curve from the debris.
\end{abstract}

\maketitle

\thispagestyle{fancy}


\section{INTRODUCTION\label{S_intro}}

A star in an orbit around a massive black hole can get tidally
disrupted during its close passage by the black hole. After several
orbital periods the debris from the disrupted star settles into an
accretion disk and gradually falls into the black hole
\cite{Rees,CLG,SC,LU}. As material gets swallowed by the black hole
intense UV or soft-X ray radiation is expected to emerge from the
innermost rings of the accretion disk
\cite{FR,LS,Frank,Phinney,SW,MT,SU,Donley}. For black hole masses $M_{bh} <
10^{7}\Msun$, tidal disruption theory predicts flares with
luminosities of the order of the Eddington luminosity with durations
of the order of months, and with spectra that peak in the UV/X-ray
band \cite{Rees,EK,Ulmer,KPL,Gezari02}. High-energy flares from the
central source illuminate the debris, the photons get absorbed, and
some are re-emitted in the optical part of the spectrum (i.e. the
light is ``reprocessed''). One of the spectral lines in which this
phenomenon can be observed is the Balmer series H$\alpha$ line
($\lambda_{rest}=6563$~\AA).

The disruption of a star begins when the star approaches the tidal
radius, $r_{t}\simeq r_{\star}(M_{bh}/M_{\star})^{1/3}$, the point
where the surface gravity of the star equals the tidal acceleration
from the black hole across the diameter of the star ($r_{\star}$ and
$M_{\star}$ are the radius and mass of the star and $M_{bh}$ is the
mass of the black hole). A $10^{6}\Msun$ black hole is often used as a
prototypical example in tidal disruption calculations. This choice is
motivated by the criterion for a solar-type star to be disrupted
before it crosses the black hole event horizon (i.e. the Schwarzchild
radius, $r_{s}$) in order for emission to be observable. For
supermassive black holes with $M_{bh} > 10^{8}\Msun$, $r_{s}>r_{t}$
and the star falls into the black hole before it gets disrupted.

\section{CALCULATION\label{S_calc}}

Tidal disruption simulations were carried out with a three dimensional
relativistic smoothed particle hydrodynamics (hereafter SPH) code in
order to study the dynamical evolution of the post-disruption
debris. The SPH code used provides a description of relativistic fluid
flows in a static curved spacetime geometry
\cite{Laguna1,Laguna}. We use it to simulate the tidal disruption of a
star in the potential of a Schwarzchild black hole.

\begin{figure}[h]
\includegraphics[width=60mm]{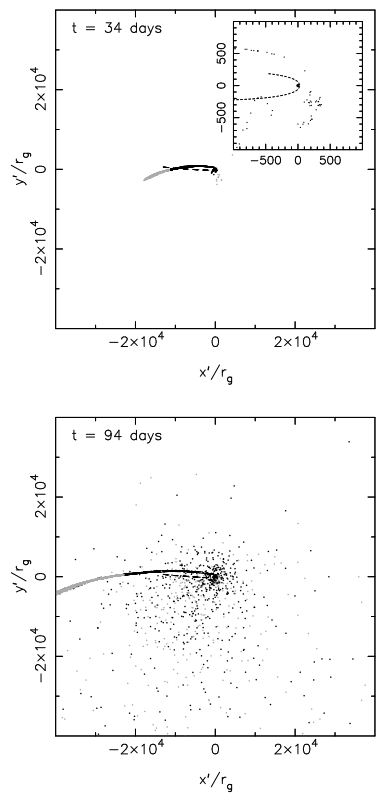}
\caption{Maps showing the positions of the SPH particles
from the 5k simulation after the second pericentric passage, projected
in the x$'$y$'$-plane at two different times. Particles
gravitationally bound to the black hole are colored black, while
unbound particles are colored grey. The dashed line represents the
initial trajectory of the star before disruption and the trajectory of
the center of mass of the debris after disruption. The maximum
particle velocities are of order $10^{-2}c$. {\it Upper Panel:}
Particle map at the start of the accretion phase, 34 days after the
disruption occurred. Inset: Particles in the inner region of the
debris, orbiting close to the black hole.{\it Lower Panel:} Particle
map at the end of the simulation, 94 days after the disruption. The
tidal tail can be clearly separated into particles that are unbound
and about to escape the black hole and particles that are returning
towards the black hole. The inner region of the debris consists of
returning particles from the inner tail that have been scattered and
form a flat rotating structure around the black hole.} \label{fig_map}
\end{figure}

Two different simulations were carried out with 5,000 and 20,000
particles (hereafter 5k and 20k respectively) contributing equally to
the mass of a $1\Msun$ star. The 5k calculation follows the debris for
94 days in total. After 34 days, significant accretion onto the black
hole begins. Our investigation follows the evolution of the line
profiles in the last 60 days. The 20k simulation spans 53 days, during
which the evolution of the line profiles was followed for the last 6
days. Using both the 5k and 20k simulations in the line profile
modeling we take advantage of the longer time span in the former and
better resolution achieved with the larger number of particles in the
latter. Figure~\ref{fig_map} shows particle distribution maps after
the second pericentric passage, at the beginning of the accretion
phase and at the end of the 5k simulation. At the early stages of the
tidal event most of the particles were located in the pronounced {\it
tidal tail}. Sixty days later, about $20\%$ of the particles are
scattered from the tidal tail and form a quasi-spherical distribution,
with most of its mass concentrated in the equatorial plane. We refer
to the spheroidal part of the debris as the {\it halo} and to its
planar component as the {\it disk}. There is a concern that some
fraction of particles of the halo are a numerical artifact of the
simulation. We further mention the implications of existance of the
spheroidal halo for the total H$\alpha$ luminosity and emission line
profiles in \S\ref{S_lcurves} and \S\ref{S_line}.

We follow the line profile calculations carried out by \cite{CH} and
\cite{ELHS} to obtain the observed profile from a Keplerian,
relativistic, thin disk in the weak field approximation. The main
objective of the calculation is to obtain the final expression for the
flux density in the observer's frame as a function of parameters
defined in the reference frame of the debris.

In order to calculate the resulting H$\alpha$ luminosity of the debris
it is necessary to determine the efficiency with which the debris
reprocesses the incident radiation. This efficiency can be
characterized by the surface emissivity of the debris as a function of
radius, $\epsilon=\epsilon_{0}\,\xi^{-q}$, where $\epsilon_{0}$ is a
constant and $\xi\equiv r/r_{\rm g}$ is the dimensionless radius
($r_{g}=r_{s}/2=G M_{bh}/c^{2}=M_{bh}$). We have used the
photoionization code CLOUDY \cite{Ferland} for numerical calculations
of the response of the debris to illumination. For a more detailed
description of the calculation refer to \cite{Bogdanovic04}.

\section{RESULTS\label{S_res}}

\subsection{Light Curves\label{S_lcurves}}

With the above emissivity prescriptions we have calculated the {\it
observed} H$\alpha$ luminosity curve of the debris at a particular
time step by computing the time at which the light was emitted from
the debris and by finding the ionizing flux that was illuminating that
location at the time the emission occurred, according to the
light-travel time from the black hole to that particular region of the
debris. We assumed that the observer is located in the x$'$z$'$ plane,
above the orbital plane at $i=30^{\circ}$ to the z$'$-axis, at a
distance $d\to +\infty$.

\begin{figure}[h]
\includegraphics[width=55mm]{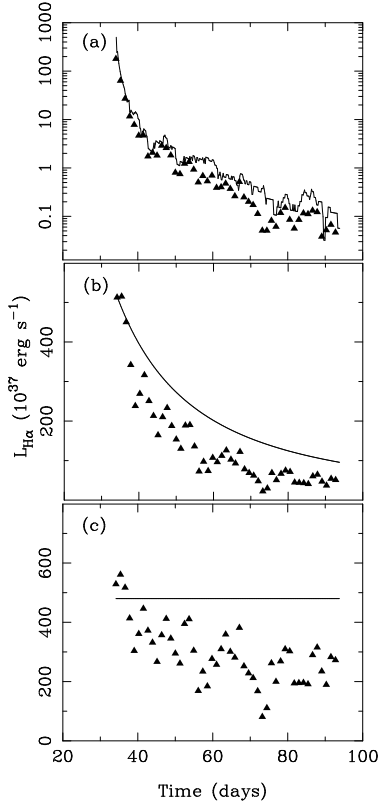}
\caption{The H$\alpha$ light curves ({\it triangles})
resulting from the reprocessing of three different illumination light
curves. The solid line shows a scaled version of the UV/X-ray
continuum light curve that illuminates the debris, which in (a) follows
the SPH accretion rate, (b) decays with time as $t^{-5/3}$, and (c) is
constant. Note that the vertical axis is calibrated logarithmically in
(a), while in (b) and (c) they have the same linear scale. The
H$\alpha$ light curves rise initially as the illumination front
propagates through the debris and then decay faster than the UV/X-ray
light curves.}\label{fig_lcurve}
\end{figure}

Figure~\ref{fig_lcurve} shows three different H$\alpha$ light curves
from the debris confined to a plane (assuming $\xi_{in}=500$,
$\xi_{out}=40,000$) during the 60-day accretion phase of the 5k
simulation.  Figure~\ref{fig_lcurve}a shows the accretion luminosity
on a logarithmic scale (solid curve), calculated from the accretion
rate of the debris in the simulation. The UV/X-ray luminosity curve is
arbitrarily scaled and overplotted on the top of the H$\alpha$ curve
for comparison. The H$\alpha$ light curve departs from the accretion
light curve at late times, though the departure appears small in the
logarithmic plot (used here due to the large dynamic range of the
light curves). The same effect is more noticeable in
Figure~\ref{fig_lcurve}b, where the accretion luminosity is
proportional to $t^{-5/3}$ and the H$\alpha$ light curve is plotted on
a linear scale. The H$\alpha$ light curve roughly follows the shape of
the incident UV/X-ray light curve at early times but decays faster at
late times. The faster decay in the H$\alpha$ light curve reflects the
debris evolution in time: as the tail becomes more elongated, the
incident photons travel a longer way to illuminate the
debris. Consequently, the intensity of the illuminating light gets
lower in the later stages of the tidal disruption event. At late times
the decay of H$\alpha$ light curve stops because of the return of more
particles from the tail to the immediate vicinity of the black hole.
As particles diffuse from the high density tail to the lower density
disk, in later stages of the simulation, their emission efficiency
increases and they contribute a significant amount of H$\alpha$ light
to the light curve. To isolate the effect of the debris evolution in
time from the evolution of the illuminating light curve, we calculate
the H$\alpha$ light curve in the case of constant illumination
(Fig.~\ref{fig_lcurve}c). Here, the relative departure of the
H$\alpha$ light curve from the UV/X-ray light curve can be interpreted
as a consequence of the expansion and redistribution of the
debris. The H$\alpha$ luminosity appears to level off at late times
because the debris disk begins to settle into a quasi-steady
configuration.

The {\it observed} H$\alpha$ flux depends sensitively on the UV/X-ray
light curve, on the distribution of matter that makes up the inner
portion of the debris, and on how quickly particles redistribute
themselves in phase space. The main features of the H$\alpha$ light
curve are: an initial rise followed by a decline, with superposed
fluctuations. The initial rise is a consequence of the propagation of
the initial illumination front through the debris. The fluctuations
are a result of the fluctuations in the accretion rate, which are
caused, in turn, by the finite number of particles employed in the
simulation. The decay rate of the H$\alpha$ light curve is determined
by the decay rate of the UV/X-ray light curve, debris expansion and
redistribution rate.

The CLOUDY calculation predicts a time-average H$\alpha$ luminosity
from the tidal debris of about $10^{36}$, $10^{37}$, and $6\times
10^{38}~{\rm erg~s^{-1}}$ for the tail, disk and halo,
respectively. In the earlier stages of the disruption event when the
UV/X-ray luminosity is super-Eddington, the H$\alpha$ luminosity is
expected to be up to 80 times higher than its average value and
comparable to that of tidal disruption candidates observed in the
local universe (see Figure~\ref{fig_lcurve}). These examples include
NGC~4450 (at 16.8~Mpc) with an H$\alpha$ luminosity of
$L_{H\alpha}$=1.8$\times 10^{39}~{\rm erg~s^{-1}}$ \cite{Ho} and
NGC~1097 (at 22~Mpc) with $L_{H\alpha}$=7.7$\times 10^{39}~{\rm
erg~s^{-1}}$ \cite{SB95}. Thus the emission-line signature of a tidal
disruption event should be detectable at least out to the distance of
the Virgo cluster. In practice, however, the detection of such
emission lines from low luminosity sources may be complicated by their
weak contrast relative to the underlying stellar continuum.

\subsection{Line Profiles from the Debris\label{S_line}}

\begin{figure}[h]
\includegraphics[width=85mm]{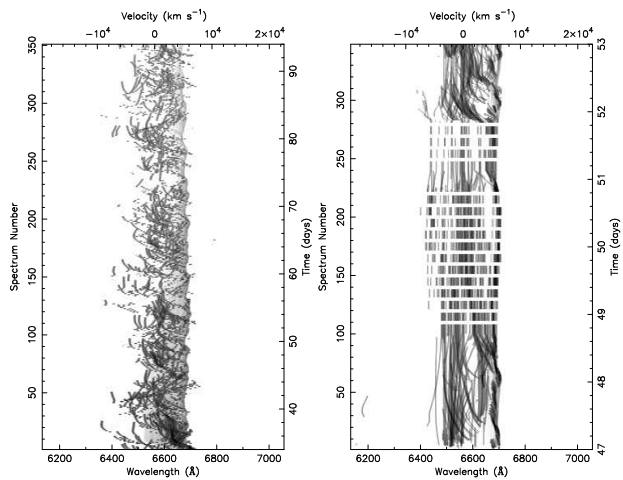}
\caption{Trailed spectrogram of the simulated H$\alpha$
emission-line profiles from the 5k simulation spanning 60 days (left)
and from the 20k simulation spanning 6 days (right). This is
effectively a 2-dimensional intensity map versus projected velocity of
the emitting material and time. Darker shades correspond to higher
intensity. The scale on the right represents time since the tidal
disruption event.}\label{fig_trail}
\end{figure}

\begin{figure}[h]
\centering
\includegraphics[width=85mm]{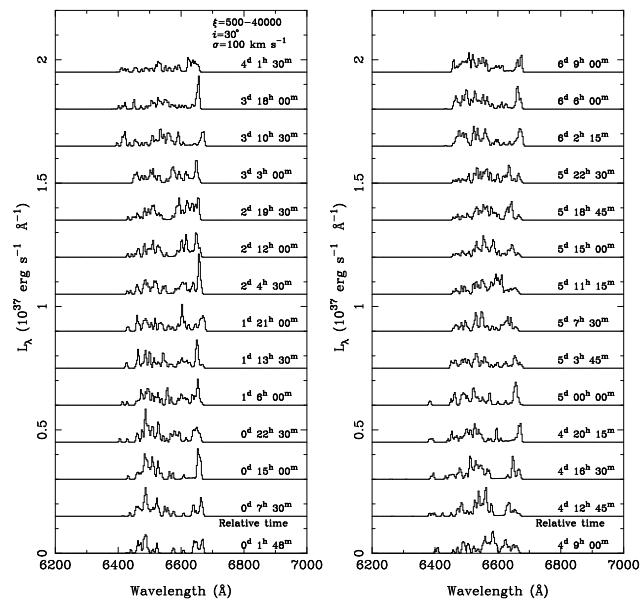}
\caption{Sequence of H$\alpha$ profiles emitted from the
region $\xi\in(500,40\,000)$ over a period of 6 days (20k run). The
relative time from the beginning of the accretion phase onto the black
hole is marked next to each profile. The accretion phase begins 47
days after the tidal disruption. The inclination of the debris plane
and the velocity shear are as marked on the figure.}\label{fig_evol}
\end{figure}

\begin{figure}[h]
\includegraphics[width=40mm]{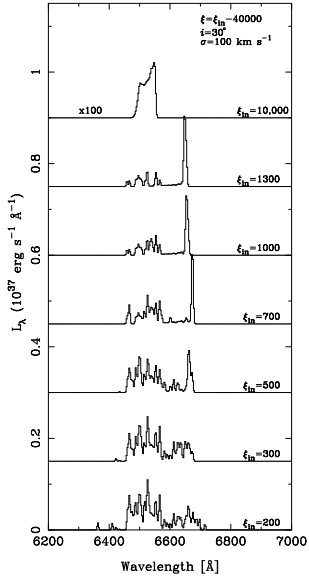}
\caption{H$\alpha$ emission line profiles simulated for seven 
different values of inner radius ($\xi_{in}$), as determined by the
propagation of the ionization front through the debris. The relative
time for profile frames is 6$^{d}$ 6$^{h}$ 0$^{m}$. The intensity of
the profile calculated for $\xi_{in}=10,000$ is multiplied by the
factor of $100$. The inclination and velocity shear are as marked on
the top of the figure.}\label{fig_rad}
\end{figure}

\begin{figure}[h]
\includegraphics[width=40mm]{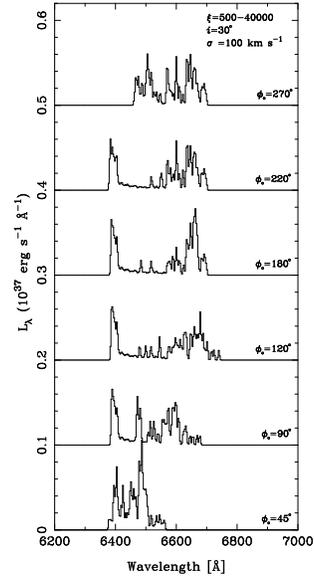}
\caption{H$\alpha$ emission line profiles simulated for six 
different azimuthal orientations of the debris with respect to the
observer, as represented by $\phi_{0}$. See profile in
Figure~\ref{fig_evol} with the time label 5$^{d}$ 22$^{h}$ 30$^{m}$
for orientation $\phi_{0}=0^{\circ}$. The size of the emitting region,
inclination and velocity shear are as marked on the top of the
figure.}\label{fig_ori}
\end{figure}

In Figures~\ref{fig_trail}--\ref{fig_ori} we show sample line profiles
to illustrate how they evolve in time and how they are affected by the
choice of model parameters and by the orientation of the observer.
Figure~\ref{fig_trail} is a ``trailed spectrogram'' summarizing the
temporal evolution of the line profiles from the two different SPH
runs; it is a 2-dimensional map of the H$\alpha$ emission as a
function of projected velocity and time. Figure~\ref{fig_evol} shows a
different representation of the evolution of the line profile with
time, which effectively comprises selected time slices from the
trailed spectrogram. Figures~\ref{fig_rad} and \ref{fig_ori} show how
the inner radius of the line-emitting region and the azimuthal
orientation of the observer affect the observed line profiles. 

A property that is immediately obvious in the line sequence is the
change of the profile shape with time 

(Figures~\ref{fig_trail} and
\ref{fig_evol}). The adopted low value of velocity dispersion allows us 
to resolve individual particles in the trailed spectrograms, orbiting
around the black hole. The evolution of the line intensities in time
roughly follows the behavior of the UV/X-ray luminosity but decays
somewhat faster in time. The multi-peaked line profile is a
consequence of the velocity field of the inner debris, which consists
of the inner portion of the tidal tail that is falling towards the
black hole (towards the observer) and debris that is rotating around
the black hole after being scattered.  The line profiles and their
variability could be observationally important features of the debris
just formed from tidal disruption. The variable line profiles might be
observed and recognized on the relatively short time scale of hours to
days.

The profiles become broader as the inner radius of the line-emitting
regions decreases since higher-velocity gas resides at smaller radii
(see Figure~\ref{fig_rad}). The approximate full width at zero
intensity of the profiles ranges from $4,500~{\rm km~s}^{-1}$ for
$\xi_{in}=10,000$ to $18,000~{\rm km~s}^{-1}$ for $\xi_{in}=200$. We
find that line profiles change from the profiles dominated by the
emission red-ward from the rest wavelength for $\xi_{in}<1000$ to
narrower profiles dominated by the blue-ward emission from the tail
for $\xi_{in}>1000$, since for large values of $\xi_{in}$, the
high-velocity rotating gas in the vicinity of the black hole is
excluded and the dominant contributions to the line profile come from
the tidal tail. The intensity of the line also decreases with
increasing inner radius, making the outer regions of the debris harder
to observe.

Because of the non-axisymmetric geometry and velocity field, the line
profiles emitted by the debris, depend on the orientation of the tidal
tail relative to the observer. In Figure~\ref{fig_ori} we show the
effect of azimuthal orientation $\phi_{0}$ of the debris, with respect
to the observer. The values of $\phi_{0}$ are $45^{\circ}$,
$90^{\circ}$, $120^{\circ}$, $180^{\circ}$, $220^{\circ}$ and
$270^{\circ}$, as measured in a counterclockwise direction from
positive x$'$-axis to the observer's line of sight. These can be
compared with the profile corresponding to the same time in
Figure~\ref{fig_evol} for $\phi_{0}=0^{\circ}$. The position of the
peaks in Figure~\ref{fig_ori} varies relative to the rest wavelength,
since the relative direction of bulk motion of the material depends on
the observer's orientation. For example, it is possible to distinguish
the emission from the tail for the range of azimuthal orientations
$90^{\circ}-220^{\circ}$. The tail emission in these profiles appears
as the most blueshifted peak, since these are the orientations for
which different portions of the tail flow towards the observer.

We have computed model profiles for several different X-ray
illumination light curves keeping all the other parameters fixed. We
used (a) the light curve obtained from the accretion rate in the 5k
simulation, (b) the light curve from the accretion rate as predicted
by theory, i.e.  $\propto t^{-5/3}$ \cite{Rees, Phinney}, and (c) a
light curve that is constant in time (Figure~\ref{fig_lcurve}). We
find that the line profile shapes do not depend sensitively on the
shape of the light curve. This is a consequence of the centrally
''weighted'' emissivity profile of the debris which causes the
innermost emission region to be the dominant contributor of the
H$\alpha$ light. In the innermost region of the debris the dynamic
range in light-travel times is not large; therefore the illumination
of the innermost emitting region is almost instantaneous. Over the
very short light-crossing time of the central emitting region, the
gradient in the UV/X-ray light curve is small and the illumination is
nearly constant over this region. The fast fluctuations in the
illuminating light curve on the other hand are smoothed out during
reprocessing in the debris, and cannot be identified in the H$\alpha$
light curve.

The temporal variability of the H$\alpha$ emission line profiles from
the post-disruption debris is one of the important indicators of a
tidal disruption event. In order to capture the rapid profile
variability, due to the variable illumination, the exposure time
should be comparable to the light crossing time of the innermost
regions of the line-emitting debris, which has the fastest and
strongest response to the ionizing radiation. Longer exposures are
expected to capture the average shape of the rapidly varying line
profiles. The light-crossing time of the innermost regions of the
debris is about $8\;\xi_{2}\;M_{6}$~minutes (where $\xi_{2}=\xi/100$
and $M_6=M_{bh}/10^6\,{\rm M_{\odot}}$), while the exposure times are
typically about 30-60 minutes (for galaxies at the distance of the
Virgo cluster, for example). Thus, if an event is caught early in its
evolution and the light-crossing time is relatively long (i.e., $M
\gtrsim 10^{6}~\Msun$), there is a chance of detecting variability
caused by changing illumination over the course of one to a few
nights. On longer time scales, variability is caused by changes in the
structure of the debris. In the presence of the spheroidal halo, the
variability of the lines may be modified by the long diffusion time
scale of photons through the halo. The component of the tidal tail
outside the halo will then still respond to the variability but on the
time scale set by the light reprocessed by the halo.

\section{CONCLUSIONS\label{S_conc}}

We modeled the emission-line luminosity and profile from the debris
released by the tidal disruption of a star by a black hole in the
early phase of evolution. Our model predicts prompt optical evolution
of the post-disruption debris and profile shapes different from
circular and elliptical disk model profiles. Since line profiles
observed so far in LINERs (low-ionization nuclear emission regions)
look more disk-like and evolve slowly, the observations are likely to
have caught the event at late times ($\ge$ 6 months after the initial
disruption), after the debris has settled into a quasi-stable
configuration.

The line profiles can take a variety of shapes for different
orientations of the debris tail relative to the observer. Due to the
very diverse morphology of the debris, it is almost impossible to
uniquely match the multi-peaked profile with the exact emission
geometry. Nevertheless, the profile widths and shifts are strongly
indicative of the velocity distribution and the location of matter
emitting the bulk of the H$\alpha$ light. Profile shapes do not depend
sensitively on the shape of the light curve of the X-rays illuminating
the debris. They strongly depend on the distance of the emitting
material from the central ionizing source, which is a consequence of
the finite propagation time of the ionization front and the
redistribution of the debris in phase space. It may be possible to
distinguish between the two effects observationally, based on their
different characteristic time scales.

If X-ray flares and the predicted variable profiles could be observed
from the same object they could be used to identify the tidal
disruption event in its early phase. The X-ray flares can be promptly
detected by all-sky synoptic X-ray surveys and high energy burst alert
missions such as {\it Swift}. The evolution of the tidal event may
then be followed with optical telescopes from the ground on longer
time scales and give an insight in the next stage of development of
the debris. Thus, simulations of the tidal disruption process on
longer time scales (of order several months to a few years) are sorely
needed. Calculations of the long-term evolution of a tidal disruption
event can predict the type of structure that the debris finally
settles into and whether its emission-line signature resembles the
transient double-peaked lines observed in LINERs. This study would
provide an important insight into the evolution of LINERs.

Finally, the observed rate of tidally disrupted solar type stars can
constrain the rate of captured compact objects which are important
gravitational wave sources \cite{SR,HB,Freitag01}, and the capture
rate of main sequence stars in our Galaxy, which are expected to emit
the peak of the gravitational radiation in the LISA frequency band and
can be detected in the local universe \cite{Freitag03}.

\bigskip 
\begin{acknowledgments}

We acknowledge the support of the Center for Gravitational Wave
Physics which is funded by the National Science Foundation under
cooperative agreement PHY 01-14375, NSF grants PHY 98-00973 and PHY
02-44788, the Zaccheus Daniel Fellowship, and the Eberly College of
Science.
\end{acknowledgments}


\end{document}